\documentclass[12pt]{article}
\usepackage{oldgerm}
\usepackage{yfonts}
\usepackage{euler}
\usepackage{graphics}
\usepackage{bm}
\usepackage{graphicx}
\usepackage{amsmath}
\usepackage{amssymb}
\usepackage{amstext}
\usepackage{amscd}
\usepackage{amsfonts}
\usepackage{appendix}
\usepackage{wasysym}
\usepackage{textcomp}
\usepackage{epstopdf}
\usepackage{color}
\DeclareMathAlphabet{\EuFrak}{U}{euf}{m}{n}
\DeclareMathAlphabet{\EuScript}{U}{eus}{m}{n}

\newcommand{\nd}{\noindent}

\title{{\bf Quantum Field Theory, Feynman-, Wheeler Propagators,
Dimensional Regularization in Configuration Space
and Convolution of Lorentz Invariant Tempered Distributions}}
\author{{A. Plastino$^{1,3,4}$, M.C.Rocca$^{1,2,3}$}, \\
\small{$^1$ Departamento de F\'{\i}sica,
Universidad Nacional de La Plata,}\\
\small{$^2$ Departamento de Matem\'{a}tica,
Universidad Nacional de La Plata,}\\
\small{$^3$ Consejo Nacional de Investigaciones Cient\'{\i}ficas
y Tecnol\'{o}gicas}\\
\small{(IFLP-CCT-CONICET)-C. C. 727, 1900 La Plata -
Argentina}\\\small{$^4$  SThAR - EPFL, Lausanne, Switzerland}}

\date{\today}

\begin{document}

\maketitle

\vspace{-5mm}

\begin{abstract}

The Dimensional Regularization of Bollini and Giambiagi  
(Phys. Lett. {\bf B 40}, 566 (1972), Il Nuovo Cim. {\bf B 12}, 20 (1972).
Phys. Rev. {\bf D 53}, 5761 (1996))
can not be defined for all Schwartz Tempered Distributions
Explicitly Lorentz Invariant (STDELI) ${\cal S}^{'}_L$.  
In this paper we overcome here such limitation
and show that it can be generalized to all
aforementioned STDELI and obtain a product in a ring
with zero divisors.

For this purpose, we resort to a formula obtained in [Int. J. of
Theor. Phys. {\bf 43}, 1019 (2004)] and  demonstrate the existence
of the convolution (in  Minkowskian space) of such
distributions. This is done by following a procedure similar to
that used so as to define a general convolution between the
Ultradistributions of J. Sebastiao e Silva [Math. Ann. {\bf 136},
38 (1958)], also known as Ultrahyperfunctions, 
obtained by Bollini et al. [Int. J. of
Theor. Phys. {\bf 38}, 2315 (1999), {\bf 43}, 1019 (2004), {\bf
43}, 59 (2004),{\bf 46}, 3030 (2007)]. Using the Inverse Fourier
Transform we get the ring with zero divisors ${\cal S}^{'}_{LA}$,
defined as ${\cal S}^{'}_{LA}={\cal F}^{-1}\{{\cal S}^{'}_L\}$,
where ${\cal F}^{-1}$ denotes the Inverse Fourier Transform. In
this manner we effect a dimensional regularization in  momentum
space (the ring ${\cal S}^{'}_{L}$) via convolution, and a product
of distributions in the corresponding configuration space (the
ring ${\cal S}^{'}_{LA})$. This generalizes the results obtained
by Bollini and Giambiagi for  Euclidean space in [Phys. Rev. {\bf
D 53}, 5761 (1996)].

We provide several examples of the application of our new
results in Quantum Field Theory. In particular, the convolution of
$n$ massless Feynman's propagators and the convolution of n massless
Wheeler's propagators in Minkowskian
space.

The results obtained in this work have already allowed us to calculate the 
classical partition function of Newtonian gravity, for the first time ever,
in the Gibbs' formulation and in the Tsallis' one:
Physica A {\bf 503}, 793 (2018), {\bf 497}, 310 (2918).

It is our hope that this convolution will allow one to quantize
non-renormalizable Quantum Fields Theories.\\ 
\nd
{\bf PACS}: 11.10.-z, 03.70.+k, 03.65.Ca, 03.65.Db.\\
\nd
{\bf KEYWORDS}:  Dimensional Regularization,
Feynman's propagators, Wheeler's propagators.

\end{abstract}

\newpage

\tableofcontents

\newpage

\renewcommand{\theequation}{\arabic{section}.\arabic{subsection}.
\arabic{equation}}

\section{Introduction}

The problem of defining the product of two distributions
(a product in a ring with divisors of zero)
is an old one of hard functional analysis.

In QFT the problem of evaluating the product
of distributions with coincident point singularities is related 
to the asymptotic behaviour of loop integrals
of propagators.

From a mathematical point of view, practically all definitions of
that product lead to limitations on the set of distributions that
can be multiplied by each other to give another distribution of
the same type.

In fact, Laurent Schwartz showed that he can not define a product
of distributions regarded as an algebra, instead of as a ring with
divisors of zero.

In  references \cite{tp3, tp18, tp19, tp20} it was demonstrated
that it is possible to define a general convolution between the
ultradistributions of JSS \cite{tp6} (Ultrahyperfunctions).
This convolution yields another Ultrahyperfunction.
Therefore, we have a product in a ring with zero divisors.
Such a ring is the space of
Distributions of Exponential Type, or Ultradistributions of
Exponential Type, obtained applying the anti-Fourier transform to
the space of Tempered Ultradistributions or Ultradistributions of
Exponential Type.

We must clarify at this point that the Ultrahyperfunctions
are the generalization and extension to the complex plane
of the Schwartz Tempered Distributions and the
Distributions of Exponential Type. That is, the
Temperate Distributions and those of Exponential Type
are a subset of the Ultrahyprefunctions.

The problem we then face  is that of formulating  the convolution
between Ultradistributions. This is a complex issue, difficult to
manage, even if  it has the advantage of allowing one to attempt
non-renormalizable QFT's.

Fortunately, we have found that a method similar to that used to
define the convolution of Ultradistributions  can also be used to
define the convolution of  Lorentz Invariant distributions using
the DR of BG in
momentum space. 
{\it As a consequence, Ultradistributions meed not to be used
in the calculations of this paper, which considerably 
simplifies it}.
Taking  advantage of such Regularization one can
also work in  configuration space \cite{tp12}. Thus, one can
obtain a convolution of Lorentz Invariant Tempered Distributions
in momentum space and the corresponding product in configuration
space.

DR is one of the most important advances in theoretical physics and is used in several disciplines of it \cite{dr1}-\cite{dr54}.

With our DR generalization happens to be
a convolution of STDELI in momentum space
and a product in a ring with divisors of zero in configuration space.

It is our hope that this convolution can then be used to treat
non-renormalizable QFT's. This is our present
goals.

More to the point let us emphasize that our work is concerned with deeper issues than those regarding QFT axiomatics in Euclidean space and QFT renormalization. 
Here, we are {\it generalizing} BG dimensional regularization   
to {\it all Schwartz Tempered, explicitly 
Lorentz invariant,  distributions
(STDELI)}, something that BG were unable to achieve. This would permit 
one to deal with non-renormalizable QFT's.
Indeed, we do not have to  use counterterms in a renormalization process devoted to eliminate infinities. 
{\it This is exactly what we do not want to do, since a non-renormalizable theory involves an infinite numeber of counterterms.} The central purpose of our work is to define a STDELI  convolution in order to avoid counterterms.
 We do not appeal to a simple correlation-functions' convolution 
 (not defined for all STDELI). At the same time, we conserve 
 all extant solutions to the problem of running coupling constants
and the renormalization group. The STDELI convolution,  once obtained, converts configuration space into a ring with zero-divisors. In it,   one has now defined a product between the ring-elements. Thus, any unitary-causal-Lorentz invariant  theory quantified in such a manner
 becomes  predictive. The distinction those between renormalizable on not-renormalizable QFT's becomes unnecessary now. 
With our BG generalization, that uses Laurent's expansions in the dimension,
all finite constants of the  convolutions become completely determined, eliminating arbitrary choices of finite constants. This is tantamount to eliminating  all 
finite renormalizations of the theory. 
What is the importance of using only {\it that term
 independent of the dimension} 
in Laurent's expansion? That the result obtained  for finite convolutions will 
coincide with such a term. This translates to configuration  space
the product-operation  in a ring
with divisors of zero.

As examples, we calculate some convolutions of distributions used
in QFT. In particular, the convolution of n
massless Feynman's propagators and the convolution of n massless
Wheeler's propagators. For a full discussion about definition and
properties of Wheeler's propagators see \cite{w1,w2} which in turn
are based on Wheeler and Feynman works \cite{w3,w4}.

The results obtained in this work have already allowed us to calculate
the classical partition function of Newtonian gravity, for the first time ever,
in the Gibbs' formulation and in the Tsallis' one \cite{aa}

Note that we have added an Appendix the by recourse to a simple example,
is able to make it explicit how much simpler is our treatment is compared
to the habitual DR technique.

\section{Preliminary Materials}

\subsection{Lorentz Invariant Tempered Distributions}

\setcounter{equation}{0}

In this subsection we give the definitions that we will use
in this paper.

We consider first the case on the $\nu$-dimensional
Minkowskian space $\boldsymbol{M}_{\nu}$
Let $\boldsymbol{S}^{'}$ be the space of Schwartz Tempered
Distributions \cite{tp6,tp7}. Let be $g\in\boldsymbol{S}^{'}$.
We say that
$g\in\boldsymbol{S}^{'}_L$ if and only if:
\begin{equation}
\label{ep2.1}
g(\rho)=\frac {d^l} {d\rho^l}f(\rho)
\end{equation}
where the derivative is in the sense of distributions,
{\it $l$ is a natural number},
$\rho=k^2=k_0^2-k_1^2-k_2^2-\cdot\cdot\cdot-k_{\nu-1}^2$,
$f$ satisfies:
\begin{equation}
\label{ep2.2}
\int\limits_{-\infty}^{\infty} \frac
{\left|f(\rho)\right|} {(1+\rho^2)^n}d\rho<\infty,
\end{equation}
and is continuous in $\boldsymbol{M}_{\nu}$. {\it The exponent $n$ is
a natural number.}
We say then that $f\in \boldsymbol{T}_{1L}$.

In the case of Euclidean space $\boldsymbol{R}_{\nu}$, let
$g\in\boldsymbol{S}^{'}$. We say that $g\in\boldsymbol{S}^{'}_R$
if and only if
\begin{equation}
\label{ep2.3} 
g(k)=\frac {d^l} {dk^l}f(k),
\end{equation}
where $k^2=k_0^2+k_1^2+k_2^2-\cdot\cdot\cdot+k_{\nu-1}^2,$ with
$f(k)$ satisfying:
\begin{equation}
\label{ep2.4} 
\int\limits_0^{\infty} \frac {\left|f(k)\right|}
{(1+k^2)^n}dk<\infty,
\end{equation}
and $f(k)$ is continuous in $\boldsymbol{R}_{\nu}$. We say then
that $f\in\boldsymbol{T}_{1R}$.

We call $\boldsymbol{S}^{'}_{LA}$ and $\boldsymbol{S}^{'}_{RA}$
the Fourier Anti-transformed Spaces of $\boldsymbol{S}^{'}_{L}$
and $\boldsymbol{S}^{'}_{R}$, respectively.

\subsection{The Fourier Transform in Euclidean Space}

\setcounter{equation}{0}

The Fourier transform of a spherically symmetric function is
given, according to Bochner's formula, by \cite{tp17}:
\begin{equation}
\label{ep3.1} 
f(k,\nu)=\frac {(2\pi)^{\frac {\nu} {2}}} {k^{\frac
{\nu-2} {2}}}\int\limits_0^{\infty} \hat{f}(r,\nu) r^{\frac {\nu}
{2}}\boldsymbol{\cal{J}}_{\frac {\nu-2} {2}}(kr)\;dr,
\end{equation}
where $r^2=x_0^2+x_1^2+\cdot\cdot\cdot+x_{\nu-1}^2$\; ; \;
$k^2=k_0^2+k_1^2+\cdot\cdot\cdot+k_{\nu-1}^2$ and
$\boldsymbol{\cal{J}}_{\frac {\nu-2} {2}}$ is the
Bessel function of order $(\nu-2)/2$.
By the use of the equality
\begin{equation}
\label{ep3.2} 
\pi\boldsymbol{\cal{J}}_{\frac {\nu-2} {2}}(z)=
e^{-i\frac {\pi} {4}\nu}\boldsymbol{\cal{K}}_{\frac {\nu-2}
{2}}(-iz)+ e^{i\frac {\pi} {4}\nu}\boldsymbol{\cal{K}}_{\frac
{\nu-2} {2}}(iz),
\end{equation}
where $\boldsymbol{\cal{K}}$ is the modified Bessel function,
(\ref{ep3.1}) takes the form:
\[f(k,\nu)=2\frac {(2\pi)^{\frac {\nu-2} {2}}} {k^{\frac {\nu-2} {2}}}\int\limits_0^{\infty}
\hat{f}(r,\nu)
r^{\frac {\nu} {2}}\left[
e^{-i\frac {\pi} {4}\nu}\boldsymbol{\cal{K}}_{\frac {\nu-2} {2}}(-ikr)+\right.\]
\begin{equation}
\label{ep3.3} 
\left. e^{i\frac {\pi}
{4}\nu}\boldsymbol{\cal{K}}_{\frac {\nu-2} {2}}(ikr)\right]\;dr.
\end{equation}
By performing the change of variables $x=r^2$,  $\rho=k^2$,
(\ref{ep3.1}) and (\ref{ep3.3})  can be re-written as:
\begin{equation}
\label{ep3.4}
f(\rho,\nu)=\pi\frac {(2\pi)^{\frac {\nu-2} {2}}} {\rho^{\frac {\nu-2} {4}}}\int\limits_0^{\infty}
\hat{f}(x,\nu)
x^{\frac {\nu-2} {4}}\boldsymbol{\cal{J}}_{\frac {\nu-2} {2}}({\rho}^{1/2}x^{1/2})\;dx
\end{equation}
\[f(\rho,\nu)=\frac {(2\pi)^{\frac {\nu-2} {2}}} {\rho^{\frac {\nu-2} {4}}}\int\limits_0^{\infty}
\hat{f}(x,\nu)
x^{\frac {\nu-2} {4}}\left[
e^{-i\frac {\pi} {4}\nu}\boldsymbol{\cal{K}}_{\frac {\nu-2} {2}}(-ix^{1/2}{\rho}^{1/2})+\right.\]
\begin{equation}
\label{ep3.5} 
\left. e^{i\frac {\pi}
{4}\nu}\boldsymbol{\cal{K}}_{\frac {\nu-2}
{2}}(ix^{1/2}{\rho}^{1/2})\right]\;dx.
\end{equation}

\subsection{The Fourier Transform in Minkowskian Space}

\setcounter{equation}{0}

For the Minkowskian case we have, according to ref.\cite{tp18}
\[f(\rho,\nu)=(2\pi)^{\frac {\nu-2} {2}}\int\limits_{-\infty}^{\infty}\hat{f}(x,\nu)\left\{
e^{\frac {i\pi(\nu-2)} {4}} \frac {(x+i0)^{\frac {\nu-2} {4}}} {(\rho+i0)^{\frac {\nu-2} {4}}}
{\cal{K}}_{\frac {\nu-2} {2}}[-i(x+i0)^{1/2}(\rho+i0)^{1/2}]\right.+\]
\begin{equation}
\label{ep4.1} 
+\left.e^{\frac {i\pi(2-\nu)} {4}} \frac
{(x-i0)^{\frac {\nu-2} {4}}} {(\rho-i0)^{\frac {\nu-2} {4}}}
{\cal{K}}_{\frac {\nu-2}
{2}}[i(x-i0)^{1/2}(\rho-i0)^{1/2}]\right\}\;dx,
\end{equation}
where $\rho=k_0^2-k_1^2-k_2^2\cdot\cdot\cdot-k_{\nu-1}^2$. The
corresponding inversion formula is then given by \cite{tp18}:
\[\hat{f}(x,\nu)=\frac {1} {(2\pi)^{\frac {\nu+2} {2}}}\int\limits_{-\infty}^{\infty}
f(\rho,\nu)\left\{
e^{\frac {i\pi(\nu-2)} {4}} \frac {(\rho+i0)^{\frac {\nu-2} {4}}} {(x+i0)^{\frac {\nu-2} {4}}}
{\cal{K}}_{\frac {\nu-2} {2}}[-i(x+i0)^{1/2}(\rho+i0)^{1/2}]\right.+\]
\begin{equation}
\label{ep4.2} 
+\left.e^{\frac {i\pi(2-\nu)} {4}} \frac
{(\rho-i0)^{\frac {\nu-2} {4}}} {(x-i0)^{\frac {\nu-2} {4}}}
{\cal{K}}_{\frac {\nu-2}
{2}}[i(x-i0)^{1/2}(\rho-i0)^{1/2}]\right\}\;d\rho.
\end{equation}
Eq. (\ref{ep4.1}) is the generalization of Bochner's formula
(\ref{ep3.1}) to  Minkowskian Space.

\setcounter{equation}{0}

\subsection{An original example}

As an example not previously published 
of  this formula we will calculate the Fourier
anti-transform of the Dirac's delta $\delta(\rho)$ in four dimensions.
For this, we make use of the formula given in \cite{tp13}:
\begin{equation}
\label{eq4.3} 
{\cal K}_1(z)=-\frac {1}
{z}+\sum\limits_{k=0}^{\infty} \frac {\left(\frac {z}
{2}\right)^{(1+2k)}} {k!(1+k)!} \left[\ln\left(\frac {z}
{2}\right)-\frac {1} {2}\psi(k+1)- \psi(k+2)\right],
\end{equation}
where $\psi(z)=\frac {d[\ln\Gamma(z)]} {dz}$. Then:
\[\hat{f}(x,4)=\frac {1} {(2\pi)^3}\int\limits_{-\infty}^{\infty}
\delta(\rho)\left\{
i\frac {(\rho+i0)^{\frac {1} {2}}} {(x+i0)^{\frac {1} {2}}}
{\cal{K}}_1[-i(x+i0)^{1/2}(\rho+i0)^{1/2}]\right.+\]
\begin{equation}
\label{ep4.4} 
-\left.i \frac {(\rho-i0)^{\frac {\nu-2} {4}}}
{(x-i0)^{\frac {\nu-2} {4}}}
{\cal{K}}_1[i(x-i0)^{1/2}(\rho-i0)^{1/2}]\right\}\;d\rho.
\end{equation}
After a simple calculation we obtain:
\begin{equation}
\label{ep4.5}
\hat{f}(x,4)=\frac {1} {(2\pi)^3}\int\limits_{-\infty}^{\infty}
\delta(\rho)\left[\frac {1} {x+i0}+\frac {1} {x-i0}\right]d\rho
\end{equation}
and finally:
\begin{equation}
\label{ep4.6} 
\hat{f}(x,4)=\frac {1} {(2\pi)^3} \left[\frac {1}
{x+i0}+\frac {1} {x-i0}\right].
\end{equation}

\section{The Convolution in Euclidean Space}

\setcounter{equation}{0}

\subsection{The generalization of Dimensional Regularization in
Configuration Space to the Euclidean Space}

The expression for the convolution of two spherically symmetric  functions
was deduced in  ref.\cite{tp12} ($h(k,\nu)=(f\ast g)(k,\nu)$):
\[h(k,\nu)=\frac {2^{4-\nu}{\pi}^{\frac {\nu-1} {2}}}
{\Gamma(\frac {\nu-1} {2})k^{\nu-2}}\iint\limits_{\;0}^{\;\;\;\infty}
f(k_1,\nu)g(k_2,\nu)\;\times\]
\begin{equation}
\label{ep5.1} 
[4k_1^2k_2^2-(k^2-k_1^2-k_2^2)^2]_+^{ \frac {\nu-3}
{2}}k_1k_2\;dk_1\;dk_2.
\end{equation}
However, BG did not obtain a
product in a ring with divisors of zero, which
we will do now.
 Consider here that $f$ and $g$ belong to $\boldsymbol{S}^{'}_R$.
With the change of variables $\rho=k^2$,  ${\rho}_1=k_1^2$,
${\rho}_2=k_2^2$ takes the form:
\[h(\rho,\nu)=\frac {2^{2-\nu}{\pi}^{\frac {\nu-1} {2}}}
{\Gamma(\frac {\nu-1} {2}){\rho}^{\frac {\nu-2} {2}}}\iint\limits_{\;0}^{\;\;\;\infty}
f({\rho}_1,\nu)g({\rho}_2,\nu)\;\times\]
\begin{equation}
\label{ep5.2} 
[4{\rho}_1{\rho}_2-(\rho-{\rho}_1-{\rho}_2)^2]_+^{
\frac {\nu-3} {2}}\;d{\rho}_1\;d{\rho}_2.
\end{equation}
Let {\textgoth{V}} be a vertical band contained in the complex
$\nu$-plane \textgoth{W}. Integral (\ref{ep5.2}) is an analytic
function of $\nu$ defined in the domain \textgoth{V}. Then,
according to the method of ref.\cite{tp3}, $h(\nu,\rho)$ can be
analytically continued to other parts of \textgoth{W}. In
particular, near the dimension $\nu_0$ we have the Laurent's
expansion:
\begin{equation}
\label{ep5.3}
h(\rho,\nu)=\sum\limits_{m=-1}^{\infty}h^{(m)}(\rho){(\nu-\nu_0)}^m.
\end{equation}
Here, $\nu_0$ is the dimension of the considered space. In
particular, $\nu_0=4$ is the dimension that we will consider. We
now define the convolution product as the
$(\nu-\nu_0)$-independent term of the Laurent's expansion.
(\ref{ep5.3}):
\begin{equation}
\label{ep5.4} 
h_{\nu_0}(\rho)=h^{(0)}(\rho).
\end{equation}
Thus, in the ring with zero divisors $\boldsymbol{S}^{'}_{RA}$, we
have defined a product of distributions.

\setcounter{equation}{0}

\subsection{Example}

As an example of the use of (\ref{ep3.1}) and (\ref{ep5.1}), we
evaluate the convolution of a massless propagator with a
propagator corresponding to a scalar particle of mass m. The
result of this convolution, using this formula, is given in
\cite{tp14}. It is:
\begin{equation}
\label{ep5.5} 
h(k,\nu)=2^{\nu-2} \pi^{\frac {\nu} {2}}m^{\nu-4}
\frac {\Gamma\left(\frac {\nu-2} {2}\right) \Gamma\left(\frac
{4-\nu} {2}\right)} {\Gamma\left(\frac {\nu} {2}\right)}
F\left(1,\frac {4-\nu} {2};\frac {\nu} {2};-\frac {k^2}
{m^2}\right).
\end{equation}
Now we use the equality:
\[\Gamma\left(\frac {4-\nu} {2}\right)
F\left(1,\frac {4-\nu} {2};\frac {\nu} {2};-\frac {k^2} {m^2}\right)=\]
\begin{equation}
\label{ep5.6} 
\Gamma\left(\frac {4-\nu} {2}\right)- \frac {2}
{\nu} \Gamma\left(\frac {6-\nu} {2}\right) \frac {k^2} {m^2}
F\left(1,\frac {6-\nu} {2};\frac {2+\nu} {2};-\frac {k^2}
{m^2}\right).
\end{equation}
After a tedious calculation, we obtain the corresponding Laurent's
expansion of $h(k,\nu)$:
\[h(k,\nu)=
-\frac {8\pi^2} {\nu-4}+4\pi^2\left(\mathbf{C}+2-\ln 4-\ln\pi-\ln m^2\right)-\]
\begin{equation}
\label{ep5.7} 
2\pi^2 \frac {k^2} {m^2} F\left(1,1;3;-\frac {k^2}
{m^2}\right)+ \sum\limits_{s=1}^{\infty}a_s(\nu-4)^s,
\end{equation}
where $\mathbf{C}$ is  Euler's constant with changed sign $\mathbf{C}=-0.57721566490$.
Thus, we have
\[\frac {1} {k^2}\ast\frac {1} {k^2+m^2}=
4\pi^2\left(\mathbf{C}+2-\ln 4-\ln\pi-\ln m^2\right)-\]
\begin{equation}
\label{ep5.8} 
2\pi^2 \frac {k^2} {m^2} F\left(1,1;3;-\frac {k^2}
{m^2}\right).
\end{equation}

\section{The Convolution in Minkowskian space}

\setcounter{equation}{0}

\subsection{The generalization of Dimensional Regularization in
Configuration Space to the Minkowskian Space}

In this section we repeat the efforts of the preceding one for
Minkowskian space.

The generalization of the Bochner's formula to  Minkowskian space
has been obtained in  reference \cite{tp18}.  The corresponding
expression for $\nu=2n$ is:
\[h(\rho,\nu)=
\frac {{\pi}^{\frac {\nu-3} {2}}} {2^{\nu-1}}
e^{\frac {i\pi(2-\nu)} {2}}
\Gamma\left(\frac {3-\nu} {2}\right)
\iint\limits_{-\infty}^{\;\;\;\infty}f({\rho}_1,\nu)g({\rho}_2,\nu)
\;\;\;\times\]
\[\left\{(\rho-i0)^{-\frac {1} {2}}
\left[\frac {(\rho-{\rho}_1-{\rho}_2)^2-4{\rho}_1{\rho}_2}
{\rho}+i0\right]^{\frac {\nu-3} {2}}+e^{i\pi(\nu-2)}\right.\;\;\;\times\]
\begin{equation}
\label{ep6.1}
\left.(\rho+i0)^{-\frac {1} {2}}
\left[\frac {(\rho-{\rho}_1-{\rho}_2)^2-4{\rho}_1{\rho}_2}
{\rho}-i0\right]^{\frac {\nu-3} {2}}\right\}d{\rho}_1\;d{\rho}_2
\end{equation}
$h(\rho,\nu)=(f\ast g)(\rho,\nu)$.\\
When $\nu=2n+1$ we obtain:
\[h(\rho,\nu)=-\frac {i{\pi}^{\frac {\nu-3} {2}}} {2^{\nu-1}
\Gamma\left(\frac {\nu-3} {2}\right)}
\iint\limits_{-\infty}^{\;\;\;\infty}f({\rho}_1,\nu)g({\rho}_2,\nu)\left[\frac{(\rho-{\rho}_1-
{\rho}_2)^2-4{\rho}_1{\rho}_2} {\rho}\right]^{\frac {\nu-3} {2}}
\left\{(\rho-i0)^{-\frac {1} {2}}
\right.\times\]
\[\left[\psi\left(\frac {\nu-1} {2}\right)+\frac {i\pi} {2}+\ln
\left[\frac{(\rho-{\rho}_1-
{\rho}_2)^2-4{\rho}_1{\rho}_2} {\rho}+i0\right]\right]-(\rho+i0)^{-\frac {1} {2}}\]
\begin{equation}
\label{ep6.2} 
\left.\left[\psi\left(\frac {\nu-1} {2}\right)+\frac
{i\pi} {2}+\ln \left[-\frac{(\rho-{\rho}_1-
{\rho}_2)^2-4{\rho}_1{\rho}_2}
{\rho}+i0\right]\right]\right\}d{\rho}_1\;d{\rho}_2.
\end{equation}
For the Minkowskian case one can also employ  Laurent's expansion
\begin{equation}
\label{ep6.3}
h(\rho,\nu)=\sum\limits_{m=-1}^{\infty}h^{(m)}(\rho){(\nu-\nu_0)}^m
\end{equation}
and therefore, again, we have for the convolution the result:
\begin{equation}
\label{ep6.4} 
h_{\nu_0}(\rho)=h^{(0)}(\rho).
\end{equation}
Thus, in the ring with zero divisors $\boldsymbol{S}^{'}_{LA}$ we
have defined a product of distributions.

\setcounter{equation}{0}

\subsection{Examples}

As an example of the use of (\ref{ep6.1}) we will consider the
convolution of two Dirac's $\delta$-distributions, $\delta(\rho)$. The result is
\[h(\rho,\nu)=
\frac {{\pi}^{\frac {\nu-3} {2}}} {2^{\nu-1}}
e^{\frac {i\pi(2-\nu)} {2}}
\Gamma\left(\frac {3-\nu} {2}\right)\]
\begin{equation}
\label{ep6.5} 
\left[(\rho-i0)^{-\frac {1} {2}}(\rho+i0)^{\frac
{\nu-3} {2}}+ e^{i\pi(\nu-2)} (\rho+i0)^{-\frac {1}
{2}}(\rho-i0)^{\frac {\nu-3} {2}} \right].
\end{equation}
Simplifying terms we obtain:
\begin{equation}
\label{ep6.6} 
h(\rho,\nu)= \frac {{\pi}^{\frac {\nu-3} {2}}}
{2^{\nu-1}} e^{\frac {i\pi(2-\nu)} {2}} \Gamma\left(\frac {3-\nu}
{2}\right) \left[\rho_+^{\frac {\nu-4} {2}}+ e^{\frac
{i\pi(\nu-2)} {2}} \rho_-^{\frac {\nu-4} {2}} \right].
\end{equation}
Thus, in four dimensions:
\begin{equation}
\label{ep6.7} 
h_4(\rho)=\delta(\rho)\ast\delta(\rho)=\frac {\pi}
{2}Sgn(\rho).
\end{equation}
Note that this convolution does not make sense in a
four-dimensional Euclidean space, since in that case
$\delta(\rho)\equiv 0$.\vskip 2mm

As a second example we calculate the convolution
$\delta(\rho-m^2)\ast\delta(\rho-m^2)$. In this case we have
\[h(\rho,\nu)=
\frac {{\pi}^{\frac {\nu-3} {2}}} {2^{\nu-1}}
e^{\frac {i\pi(2-\nu)} {2}}
\Gamma\left(\frac {3-\nu} {2}\right)\]
\begin{equation}
\label{ep6.8} 
\left[(\rho-i0)^{-\frac {1}
{2}}(\rho-2m^2+i0)^{\frac {\nu-3} {2}}+ e^{i\pi(\nu-2)}
(\rho+i0)^{-\frac {1} {2}}(\rho-2m^2-i0)^{\frac {\nu-3} {2}}
\right].
\end{equation}
When $\nu=4$ we obtain
\[\delta(\rho-m^2)\ast\delta(\rho-m^2)=\]
\begin{equation}
\label{ep6.9} 
\frac {\pi} {4} \left[(\rho-i0)^{-\frac {1}
{2}}(\rho-2m^2+i0)^{\frac {1} {2}}+ e^{i\pi(\nu-2)}
(\rho+i0)^{-\frac {1} {2}}(\rho-2m^2-i0)^{\frac {1} {2}} \right].
\end{equation}

\section{The Convolution of n massless Feynman's Propagators}

\setcounter{equation}{0}

\subsection{The Minkowskian Space Case}

Let us now calculate the convolution of n massless Feynman's
propagators ($n\geq 2$). For this purpose we take into account
that
\begin{equation}
\label{ep7.1}
{\cal F}^{-1}\left\{f_1\ast f_2\ast\cdot\cdot\cdot\ast f_n\right\}=
(2\pi)^{(n-1)\nu}\hat{f}_1\hat{f}_2\cdot\cdot\cdot\hat{f}_n
\end{equation}
According to reference \cite{tp7}, we have
\begin{equation}
\label{ep7.2} 
{\cal F}^{-1}\left\{(\rho+i0)^{-1}\right\}= \frac
{e^{-\frac {i\pi} {2}(\nu-1)}} {(2\pi)^{\nu}}
2^{(\nu-2)}\pi^{\frac {\nu} {2}} \Gamma\left(\frac {\nu}
{2}-1\right) (x-i0)^{1-\frac {\nu} {2}},
\end{equation}
and therefore,
\[{\cal F}^{-1}\left\{(\rho+i0)^{-1}\ast(\rho+i0)^{-1}\ast\cdot
\cdot\cdot\ast(\rho+i0)^{-1}\right\}=\]
\begin{equation}
\label{ep7.3} 
(2\pi)^{(n-1)\nu} \frac {e^{-\frac {i\pi}
{2}(\nu-1)n}} {(2\pi)^{\nu n}} 2^{(\nu-2)n}\pi^{\frac {\nu n} {2}}
\left[\Gamma\left(\frac {\nu} {2}-1\right)\right]^n
(x-i0)^{n(1-\frac {\nu} {2})}.
\end{equation}
Using again  reference \cite{tp7} we have now
\[{\cal F}\left\{(x-i0)^{n(1-\frac {\nu} {2})}\right\}=\]
\begin{equation}
\label{ep7.4} 
\frac {e^{-\frac {i\pi} {2}(\nu-1)}}
{\Gamma\left[n\left(\frac {\nu} {2}-1\right)\right]}
2^{\nu+2n\left(1-\frac {\nu} {2}\right)}\pi^{\frac {\nu} {2}}
\Gamma\left[\frac {\nu} {2}+n\left(1-\frac {\nu} {2}\right)\right]
(\rho+i0)^{n\left(\frac {\nu} {2}-1\right)-\frac {\nu} {2}},
\end{equation}
with which we obtain
\[(\rho+i0)^{-1}\ast(\rho+i0)^{-1}\ast\cdot
\cdot\cdot\ast(\rho+i0)^{-1}=\]
\begin{equation}
\label{ep7.5} 
\frac {e^{-\frac {i\pi} {2}(n-1)(\nu-1)}}
{\Gamma\left[n\left(\frac {\nu} {2}-1\right)\right]} \pi^{\frac
{\nu} {2}(n-1)} \left[\Gamma\left(\frac {\nu}
{2}-1\right)\right]^n \Gamma\left[\frac {\nu} {2}+n\left(1-\frac
{\nu} {2}\right)\right] (\rho+i0)^{n\left(\frac {\nu}
{2}-1\right)-\frac {\nu} {2}}.
\end{equation}
We have then, for the convolution of n massless Feynman's
propagators, the result
\[i(\rho+i0)^{-1}\ast i(\rho+i0)^{-1}\ast\cdot
\cdot\cdot\ast i(\rho+i0)^{-1}=\]
\begin{equation}
\label{ep7.6} 
\frac {e^{\frac {i\pi} {2}[n-(n-1)(\nu-1)]}}
{\Gamma\left[n\left(\frac {\nu} {2}-1\right)\right]} \pi^{\frac
{\nu} {2}(n-1)} \left[\Gamma\left(\frac {\nu}
{2}-1\right)\right]^n \Gamma\left[\frac {\nu} {2}+n\left(1-\frac
{\nu} {2}\right)\right] (\rho+i0)^{n\left(\frac {\nu}
{2}-1\right)-\frac {\nu} {2}}.\end{equation}
 After a tedious
calculation we obtain the corresponding Laurent's expansion around
$\nu=4$:
\[i(\rho+i0)^{-1}\ast i(\rho+i0)^{-1}\ast\cdot
\cdot\cdot\ast i(\rho+i0)^{-1}=
\frac {2i\pi^{2(n-1)}\rho^{n-2}}
{[\Gamma(n)]^2(\nu-4)}+\]
\[\frac {i\pi^{2(n-1)}\rho^{n-2}}
{\Gamma(n)\Gamma(n-1)}
\left[\ln(\rho+i0)-i\pi+\ln(\pi)+\frac {n} {n-1}\psi(1)-
\frac {n} {n-1}\psi(n)-\right.\]
\begin{equation}
\label{ep7.7} 
\left.\psi(n-1)\right]+\sum\limits_{m=1}^{\infty}
a_m(\rho)(\nu-4)^m.
\end{equation}
The independent $\nu-4$ term is the result of the convolution in
four dimensions
\[[i(\rho+i0)^{-1}\ast i(\rho+i0)^{-1}\ast\cdot
\cdot\cdot\ast i(\rho+i0)^{-1}]_{\nu_0=4}=\]
\[\frac {i\pi^{2(n-1)}\rho^{n-2}}
{\Gamma(n)\Gamma(n-1)}
\left[\ln(\rho+i0)-i\pi+\ln(\pi)+\frac {n} {n-1}\psi(1)-
\frac {n} {n-1}\psi(n)-\right.\]
\begin{equation}
\label{ep7.8} 
\left.\psi(n-1)\right].
\end{equation}

\setcounter{equation}{0}

\subsection{The Euclidean Space Case}

Let   us now calculate the convolution of n massless Feynman's
propagators ($n\geq 2$) in Euclidean space, using again
(\ref{ep7.1}). According to  reference \cite{tp7}, we obtain
\begin{equation}
\label{ep7.9} 
{\cal F}^{-1}\left\{k^{-2}\right\}= \frac {1}
{(2\pi)^{\nu}} 2^{(\nu-2)}\pi^{\frac {\nu} {2}} \Gamma\left(\frac
{\nu} {2}-1\right). r^{2-\nu}
\end{equation}
For n propagators we have then
\[{\cal F}^{-1}\left\{k^{-2}\ast k^{-2}\ast\cdot
\cdot\cdot\ast k^{-2}\right\}=\]
\begin{equation}
\label{ep7.10} 
\frac {(2\pi)^{(n-1)\nu}} {(2\pi)^{\nu n}}
2^{(\nu-2)n}\pi^{\frac {\nu n} {2}} \left[\Gamma\left(\frac {\nu}
{2}-1\right)\right]^n r^{n(2-\nu)}.
\end{equation}
Appealing again to reference \cite{tp7} , we can evaluate the
corresponding Fourier Transform
\[{\cal F}\left\{r^{n(2-\nu)}\right\}=\]
\begin{equation}
\label{ep7.11} 
\frac {1} {\Gamma\left[n\left(\frac {\nu}
{2}-1\right)\right]} 2^{\nu+2n\left(1-\frac {\nu}
{2}\right)}\pi^{\frac {\nu} {2}} \Gamma\left[\frac {\nu}
{2}+n\left(1-\frac {\nu} {2}\right)\right]
k^{n\left(\nu-2\right)-\nu}.
\end{equation}
Thus,
\[k^{-2}\ast k^{-2}\ast\cdot
\cdot\cdot\ast k^{-2}=\]
\begin{equation}
\label{ep7.12} 
\frac {\pi^{\frac {\nu} {2}(n-1)} }
{\Gamma\left[n\left(\frac {\nu} {2}-1\right)\right]}
\left[\Gamma\left(\frac {\nu} {2}-1\right)\right]^n
\Gamma\left[\frac {\nu} {2}+n\left(1-\frac {\nu} {2}\right)\right]
k^{n\left(\nu-2\right)-\nu}.
\end{equation}
  Let $\rho=k^2$.
We have then for the convolution of n massless Feynman's propagators
the result
\[\rho^{-1}\ast \rho^{-1}\ast\cdot
\cdot\cdot\ast \rho^{-1}=\]
\begin{equation}
\label{ep7.13} 
\frac {\pi^{\frac {\nu} {2}(n-1)}}
{\Gamma\left[n\left(\frac {\nu} {2}-1\right)\right]}
\left[\Gamma\left(\frac {\nu} {2}-1\right)\right]^n
\Gamma\left[\frac {\nu} {2}+n\left(1-\frac {\nu} {2}\right)\right]
\rho^{n\left(\frac {\nu} {2}-1\right)-\frac {\nu} {2}}.
\end{equation}
 By recourse to  Laurent's expansion we obtain
\[\rho^{-1}\ast \rho^{-1}\ast\cdot
\cdot\cdot\ast \rho^{-1}=
\frac {2{(-1)^{n-1}}\pi^{2(n-1)}\rho^{n-2}}
{[\Gamma(n)]^2(\nu-4)}+\]
\[\frac {{(-1)^{n-1}}\pi^{2(n-1)}\rho^{n-2}}
{\Gamma(n)\Gamma(n-1)}
\left[\ln(\rho)+\ln(\pi)+\frac {n} {n-1}\psi(1)-
\frac {n} {n-1}\psi(n)-\right.\]
\begin{equation}
\label{ep7.16} 
\left.\psi(n-1)\right]+\sum\limits_{m=1}^{\infty}
a_m(\rho)(\nu-4)^m.
\end{equation}
The result of the convolution in four dimensions is then
\[[\rho^{-1}\ast \rho^{-1}\ast\cdot
\cdot\cdot\ast \rho^{-1}]_{\nu_0=4}=\]
\[\frac {{(-1)^{n-1}}\pi^{2(n-1)}\rho^{n-2}}
{\Gamma(n)\Gamma(n-1)}
\left[\ln(\rho)+\ln(\pi)+\frac {n} {n-1}\psi(1)-
\frac {n} {n-1}\psi(n)-\right.\]
\begin{equation}
\label{ep7.17} 
\left.\psi(n-1)\right].
\end{equation}

We emphasize that the results of this section are
completely original.

\section{The Convolution of massless Wheeler's Propagators}

\subsection{The Convolution of two massless Wheeler's Propagators}

\setcounter{equation}{0}

The Wheeler's massless propagator is given by (note that this
propagator can not be defined in  Euclidean space)
\begin{equation}
\label{ep8.1} 
W(\rho)=\frac {i} {2}\left[\frac {1} {\rho+i0}+\frac
{1} {\rho-i0}\right],
\end{equation}
and can be written in the form:
\begin{equation}
\label{ep8.2} 
W(\rho)=\frac {i} {\rho+i0}-\pi\delta(\rho).
\end{equation}
Therefore, we have
\begin{equation}
\label{ep8.3} 
W(\rho)\ast W(\rho)=\frac {i} {\rho+i0}\ast\frac {i}
{\rho+i0} -2\pi\delta(\rho)\ast\frac {i} {\rho+i0}+
\pi^2\delta(\rho)\ast\delta(\rho).
\end{equation}
After a long and tedious calculation, using (\ref{ep6.1}) we
obtain
\[-2\pi\delta(\rho)\ast\frac {i} {\rho+i0}=
\frac {-i\pi^{\frac {\nu-1} {2}}} {2^{\nu-2}}
e^{i\pi(\frac {2-\nu} {2})}
\Gamma\left(\frac {3-\nu} {2}\right)
\Gamma(\nu-2)\Gamma(3-\nu)\times\]
\[\left\{\left[1+e^{i\pi(\nu-2)}\right]
\left[1-e^{-i\pi(3-\nu)}\right]H(\rho)\rho^{\frac {\nu} {2} -1}\right.+\]
\begin{equation}
\label{ep8.4} 
\left.2e^{i\pi(\frac {\nu-2} {2})}
\left[e^{i\pi(\nu-2)}-1\right]H(-\rho)(-\rho)^{\frac {\nu}
{2}-2}\right\}.
\end{equation}
This last equation can be re-written in the form
\[-2\pi\delta(\rho)\ast\frac {i} {\rho+i0}=
\frac {\pi^{\frac {\nu+3} {2}}e^{\frac {i\pi} {2}(3-\nu)}
\cos\pi\left(\frac {\nu-2} {2}\right)}
{2^{\nu-4}\Gamma\left(\frac {\nu-1} {2}\right)
\sin\pi\left(\frac {\nu-3} {2}\right)\sin\pi\nu}\]
\begin{equation}
\label{ep8.5} 
\left\{\cos\pi\left(\frac {\nu-2} {2}\right)
H(\rho)\rho^{\frac {\nu} {2} -1}- e^{i\pi(\nu-2)}
H(-\rho)(-\rho)^{\frac {\nu} {2}-2}\right\}.
\end{equation}
For the first convolution of (\ref{ep8.3}), we have from
(\ref{ep7.6}), with $n=2$
\begin{equation}
\label{ep8.6} 
\frac {i} {\rho+i0}\ast\frac {i} {\rho+i0}= \frac
{e^{i\frac {\pi} {2}(3-\nu)}\pi^{\frac {\nu} {2}}} {\Gamma(\nu-2)}
\left[\Gamma\left(\frac {\nu} {2}-1\right)\right]^2
\Gamma\left(2-\frac {\nu} {2}\right) (\rho+i0)^{\frac {\nu}
{2}-2}.
\end{equation}
This equation can be re-written in the form:
\begin{equation}
\label{ep8.7} 
\frac {i} {\rho+i0}\ast\frac {i} {\rho+i0}= \frac
{e^{i\frac {\pi} {2}(3-\nu)}\pi^{\frac {\nu-3} {2}}
\cos\left(\frac {\nu-3} {2}\right)} {2^\nu\Gamma\left(\frac
{\nu-1} {2}\right) \sin\pi\nu} (\rho+i0)^{\frac {\nu} {2}-2}.
\end{equation}
When $\nu=4$, the sum of (\ref{ep8.5}) and (\ref{ep8.7}) has as a
result
\begin{equation}
\label{ep8.8}
\frac {i} {\rho+i0}\ast\frac {i} {\rho+i0}
-2\pi\delta(\rho)\ast\frac {i} {\rho+i0}=
\pi^3H(-\rho)
\end{equation}
Using now (\ref{ep6.7}),  we find
\begin{equation}
\label{ep8.9} 
W(\rho)\ast W(\rho)=\frac {\pi^3} {2}.
\end{equation}

This result was obtained in the reference \cite{tp19},
formula (6.12) using the convolution of even Tempered Ultradistributions. The coincidence of (\ref{ep8.9}) with (6.12) of \cite{tp19} confirms the validity of 
the results obtained in section 6 of this paper. We emphasize 
that the present results are obtained in a manner considerably 
simpler to that of \cite{tp19}.

\subsection{The Convolution of n massless Wheeler's Propagators}

\setcounter{equation}{0}

According to reference \cite{tp7}, we have
\begin{equation}
\label{ep9.1} 
{\cal F}^{-1}\left\{(\rho+i0)^{-1}\right\}= \frac
{e^{-\frac {i\pi} {2}(\nu-1)}} {(2\pi)^{\nu}}
2^{(\nu-2)}\pi^{\frac {\nu} {2}} \Gamma\left(\frac {\nu}
{2}-1\right) (x-i0)^{1-\frac {\nu} {2}},
\end{equation}
\begin{equation}
\label{ep9.2} 
{\cal F}^{-1}\left\{(\rho-i0)^{-1}\right\}= \frac
{e^{\frac {i\pi} {2}(\nu-1)}} {(2\pi)^{\nu}}
2^{(\nu-2)}\pi^{\frac {\nu} {2}} \Gamma\left(\frac {\nu}
{2}-1\right) (x+i0)^{1-\frac {\nu} {2}},
\end{equation}
Thus,
\begin{equation}
\label{ep9.3} 
{\cal F}^{-1}\left\{W(\rho)\right\}= \frac
{i{\pi}^{\frac {\nu}  {2}}} {(2\pi)^{\nu}}
2^{(\nu-2)} \Gamma\left(\frac {\nu}
{2}-1\right) \sin\left(\frac {\pi\nu} {2}\right)
x_+^{1-\frac {\nu} {2}},
\end{equation}
As a consequence we obtain for n Wheeler's propagators
\[{\cal F}^{-1}\left\{W(\rho)\ast W(\rho)\ast\cdot\cdot\cdot 
\ast W(\rho)\right\}=\]
\begin{equation}
\label{ep9.4}
\frac {i^n{\pi}^{n\frac {\nu}  {2}}} {(2\pi)^{n\nu}}
2^{n(\nu-2)} \left[\Gamma\left(\frac {\nu}
{2}-1\right)\right]^n \sin^n\left(\frac {\pi\nu} {2}\right)
x_+^{n\left(1-\frac {\nu} {2}\right)},
\end{equation}
Resorting again to reference \cite{tp7} we have:
\[{\cal F}\left\{x_+^{n\left(1-\frac {\nu} {2}\right)}\right\}=
{\pi}^{\frac {\nu}  {2}-1} 2^{(1-n)\nu+2n}
\Gamma\left(n+1-\frac {n\nu}
{2}\right)
\Gamma\left[n-\frac {(n-1)\nu}
{2}\right]\otimes\]
\begin{equation}
\label{ep9.5}
\frac {1} {2}\left\{e^{-i\pi\left[n-(n-1)\frac {\nu} {2}\right]}
(\rho-i0)^{(n-1)\frac {\nu} {2}-n}+
e^{i\pi\left[n-(n-1)\frac {\nu} {2}\right]}
(\rho+i0)^{(n-1)\frac {\nu} {2}-n}\right\}
\end{equation}
Using (\ref{ep9.5}) we obtain finally:
\[W(\rho)\ast W(\rho)\ast\cdot\cdot\cdot 
\ast W(\rho)=\]
\[\frac {i^n{\pi}^{\frac {\nu}  {2}(n-1)-1}} {2}
\left[\Gamma\left(\frac {\nu}
{2}-1\right)\right]^n
\Gamma\left(n+1-\frac {n\nu}
{2}\right)
\Gamma\left[n-\frac {(n-1)\nu}
{2}\right]
\sin^n\left(\frac {\pi\nu} {2}\right)
\otimes\]
\begin{equation}
\label{ep9.6}
\left\{e^{i\pi\left[n-(n-1)\frac {\nu} {2}\right]}
(\rho+i0)^{(n-1)\frac {\nu} {2}-n}+
e^{-i\pi\left[n-(n-1)\frac {\nu} {2}\right]}
(\rho-i0)^{(n-1)\frac {\nu} {2}-n}\right\}
\end{equation}
We see that formula (\ref{ep9.6}) has a zero of order $n-2$ for $\nu\geq 4$, 
$\nu$ even, and 
consequently cancels for those dimensions when $n\geq 3$.
Thus we can affirm that for $\nu=4$
\begin{equation}
\label{ep9.7}
W(\rho)\ast W(\rho)\ast\cdot\cdot\cdot 
\ast W(\rho)=0
\end{equation}
when $n\geq 3$.

\newpage

\section{Discussion}

In QFT, when we use perturbative expansions, we
are dealing with products of distributions in  configuration space
or, what is the same, with convolutions of distributions in
momentum space.

In four earlier papers \cite{tp3,tp18,tp19,tp20} we have
demonstrated the existence of the convolution of JSS
Ultradistributions. This convolution allows us to treat non
renormalizable QFT's, but has the disadvantage
of being extremely complex.

Following a procedure similar to those of the previously mentioned
papers, we defined the convolution of Lorentz Invariant Temperatd
Distributions using the DR of BG.

Using this convolution we have obtained, for example, the
convolution of n massless Feynman's propagators both in Minkowskian
and Euclidean spaces and the convolution of two massless Wheeler's
propagators.

It is our hope that this convolution will allow one to treat
non-renormalizable QFT's.

\newpage

\renewcommand{\thesection}{\Alph{section}}

\renewcommand{\theequation}{\Alph{section}.\arabic{equation}}

\setcounter{section}{0}

\section{Appendix}

\setcounter{equation}{0}

The purpose of this appendix is to compare the generalization of the DR obtained
in this paper with the usual BG DR and show the differences between them. For this
we consider the convolution of two massless propagators in Euclidean space.
We start then with the usual formula for the convolution in four dimensions:
\begin{equation}
\label{a.1}
[\rho^{-1}\ast\rho^{-1}]_{\nu=4}=\int\frac {d^4p} {\vec{p}^2(\vec{p}-\vec{k})^2}
\end{equation}
The generalization of the previous convolution to $\nu$ dimensions is
\begin{equation}
\label{a.2}
\rho^{-1}\ast\rho^{-1}=\int\frac {d^\nu p} {\vec{p}^2(\vec{p}-\vec{k})^2}
\end{equation}
Using the Feynman's parameters:
\begin{equation}
\label{a.3}
\frac {1} {AB}=\int\limits_0^1\frac {dx} {[Ax+B(1-x)]^2}
\end{equation},
we can write the convolution as:
\begin{equation}
\label{a.4}
\rho^{-1}\ast\rho^{-1}=\int d^\nu p\int\limits_0^1\frac {dx}
{[(\vec{p}-\vec{k})^2x+\vec{p}^2(1-x)]^2}=
\int\limits_0^1dx\int\frac {d^\nu p}
{[(\vec{p}-\vec{k})^2x+\vec{p}^2(1-x)]^2}
\end{equation},
or more simply:
\begin{equation}
\label{a.5}
\rho^{-1}\ast\rho^{-1}=
\int\limits_0^1dx\int\frac {d^\nu p}
{[(\vec{p}-\vec{k}x)^2+\vec{k}^2x(1-x)]^2}
\end{equation}
Making the change of variable:
$\vec{s}=\vec{p}-\vec{k}x$ and calling $a=\vec{k}^2x(1-x)$ we obtain:
\begin{equation}
\label{a.6}
\rho^{-1}\ast\rho^{-1}=
\int\limits_0^1dx\int\frac {d^\nu s}
{(\vec{s}^2+a)^2}
\end{equation},
equivalently:
\begin{equation}
\label{a.7}
\rho^{-1}\ast\rho^{-1}=
\frac {2\pi^{\frac {\nu} {2}}} {\Gamma\left(\frac {\nu} {2}\right)}
\int\limits_0^1dx\int\limits_0^{\infty}\frac {s^{\nu-1}}
{(s^2+a)^2}ds
\end{equation}
Making the change of variable $y=s^2$ we have:
\begin{equation}
\label{a.8}
\rho^{-1}\ast\rho^{-1}=
\frac {\pi^{\frac {\nu} {2}}} {\Gamma\left(\frac {\nu} {2}\right)}
\int\limits_0^1dx\int\limits_0^{\infty}\frac {y^{\frac {\nu} {2}-1}}
{(y+a)^2}dy
\end{equation}
Using \cite{tp13} we can calculate the previous integral.
The result is:
\begin{equation}
\label{a.9}
\rho^{-1}\ast\rho^{-1}=
\pi^{\frac {\nu} {2}}\Gamma\left(2-\frac {\nu} {2}\right)
\int\limits_0^1a^{\frac {\nu} {2}-2}dx
\end{equation},
or in an equivalent way:
\begin{equation}
\label{a.10}
\rho^{-1}\ast\rho^{-1}=
\pi^{\frac {\nu} {2}}\Gamma\left(2-\frac {\nu} {2}\right)\rho^{\frac {\nu} {2}-2}
\int\limits_0^1[x(1-x)]^{\frac {\nu} {2}-2}dx
\end{equation}
By recourse again to the results given in \cite{tp13} we have:
\begin{equation}
\label{a.11}
\rho^{-1}\ast\rho^{-1}=
\frac {\pi^{\frac {\nu} {2}}\left[\Gamma\left(\frac {\nu} {2}-1\right)\right]^2
\rho^{\frac {\nu} {2}-2}}
{\Gamma(\nu-2)}
\Gamma\left(2-\frac {\nu} {2}\right)
\end{equation}
We notice now that (\ref{a.11}) can be written in the form:
\begin{equation}
\label{a.12}
\rho^{-1}\ast\rho^{-1}=
\frac {2\pi^2} {4-\nu}-\pi^2[\ln\rho+\ln\pi-\psi(2)]+
\sum\limits_{k=1}^{\infty}a_k(\nu-4)^k
\end{equation}
We see that the four-dimensional convolution is not univocally defined
from the $\nu$-dimensional convolution since there are several ways to choose
its finite part.

If we resort to our generalization and select the independent term
of $\nu-4$ we get for the four-dimensional convolution:
\begin{equation}
\label{a.13}
[\rho^{-1}\ast\rho^{-1}]_{\nu=4}=
-\pi^2[\ln\rho+\ln\pi-\psi(2)]
\end{equation}
which coincides for $n=2$ with our result given in (\ref{ep7.17})

We should note that the calculation made in this appendix is more complex than the
obtained with our generalization of the DR for n masless propagators.
In fact, if we wanted to evaluate the convolution of n massless
propagators with the usual method of DR used in this appendix, we would have
to perform a very long calculation that would involve a large number of integrals.
If we where able to obtain 
the correct result, the four-dimensional convolution
would not be completely determined.


\newpage

\begin{thebibliography}{99}

\bibitem{tp14} C. G. Bollini and J. J. Giambiagi: Phys. Lett.
{\bf B 40}, 566 (1972).

\bibitem{tp15} C. G. Bollini and J. J. Giambiagi:Il Nuovo Cim.
{\bf B 12}, 20 (1972).

\bibitem{tp12} C. G. Bollini and J.J Giambiagi :
Phys. Rev. {\bf D 53}, 5761 (1996).

\bibitem{tp4} L. Schwartz : ``Th\'eorie des distributions''.
Hermann, Paris (1966)

\bibitem{tp3} C. G. Bollini, T. Escobar and M. C. Rocca :
Int. J. of Theor. Phys. {\bf 38}, 2315 (1999).

\bibitem{tp18} C. G. Bollini and M. C. Rocca :
Int. J. of Theor. Phys. {\bf 43}, 1019 (2004).

\bibitem{tp19} C. G. Bollini and M. C. Rocca :
Int. J. of Theor. Phys. {\bf 43}, 59 (2004).

\bibitem{tp20} C. G. Bollini, P. Marchiano and M. C. Rocca :
Int. J. of Theor. Phys. {\bf 46}, 3030 (2007).

\bibitem{tp6} J. Sebastiao e Silva : Math. Ann. {\bf 136},
38 (1958).

\bibitem{dr1} D. Berenstein and A. Miller:
Phys. Rev. D {\bf 90}, 086011 (2014).

\bibitem{dr2} D. Anselmi:
Phys. Rev. D {\bf 89}, 125024  (2014).

\bibitem{dr3} P. Jaranowski and G. Schäfer:
Phys. Rev. D {\bf 87}, 081503(R)  (2013).

\bibitem{dr4} T. Inagaki, D. Kimura, H. Kohyama, and A. Kvinikhidze:
Phys. Rev. D {\bf 86}, 116013 (2012).

\bibitem{dr5} J. Qiu:
Phys. Rev. D {\bf 77}, 125032 (2008).

\bibitem{dr6} L. Blanchet, T. Damour, G. Esposito-Farèse, and B. R. Iyer:
Phys. Rev. D {\bf 71}, 124004 (2005).

\bibitem{dr7} F. Bastianelli, O. Corradini, and A. Zirotti:
Phys. Rev. D {\bf 67}, 104009 (2003).

\bibitem{dr8} D. Lehmann and G. Prézeau:
Phys. Rev. D {\bf 65}, 016001 (2001).

\bibitem{dr9} A. P. Baêta Scarpelli, M. Sampaio, and M. C. Nemes:
Phys. Rev. D {\bf 63}, 046004 (2001).

\bibitem{dr10} E. Braaten and Yu-Qi Chen:
Phys. Rev. D {\bf 55}, 7152 (1997).

\bibitem{dr11} J. Smith and W. L. van Neerven
EPJ C {\bf 40}, 199 (2005).

\bibitem{dr12} J. F. Schonfeld
EPJ C {\bf 76}, 710 (2016).

\bibitem{dr13} C. Gnendiger et al.:
EPJ C {\bf 77}, 471 (2017).   
    
\bibitem{dr14} P. Arnold, Han-Chih Chang and S. Iqbal: 
JHEP 100 (2016).
    
\bibitem{dr15} I. AravE, Y. Oz and A. Raviv-Moshe: 
JHEP 88 (2017).

\bibitem{dr16} C. Anastasiou, S. Buehler, C. Duhr and F. Herzog:
JHEP 62 (2012).
    
\bibitem{dr17} F. Niedermayer and P. Weisz:
JHEP 110 (2016).
    
\bibitem{dr18} C. Coriano, L. Delle Rose, E. Mottolaand M. Serino:
JHEP 147 (2012).

\bibitem{dr19} F. Dulat, S. Lionetti, B. Mistlberger,A. Pelloni and C. Specchia:
JHEP 17 (2017).
    
\bibitem{dr20} T. Gehrmann and N. Greiner:
JHEP 50 (2010).

\bibitem{dr21} T.Lappia and R.Paatelainena:
Ann. of Phys. {\bf 379}, 34 (2017).

\bibitem{dr22} S.Grooteab, J.G.Körner and A.A.Pivovarov:
Ann. of Phys. {\bf 322}, 2374 (2007).

\bibitem{dr23} N.C.Tsamis and R.P.Woodard:
Ann. of Phys. {\bf 321}, 875 (2006).

\bibitem{dr24} S. Krewaland and K. Nakayama:
Ann. of Phys. {\bf 216}, 210 (1992).

\bibitem{dr25}  L. Rosen and J. D. Wright
Comm. Math. Phys. {\bf 134}, 433 (1990).

\bibitem{dr26}  F. David
Comm. Math. Phys. {\bf 81}, 149 (1981).

\bibitem{dr27}  P. Breitenlohner and D. Maison
Comm. Math. Phys. {\bf 52}, 11 (1977).

\bibitem{dr28} S. Teber and A. V. Kotikov:
EPL {\bf 107}, 57001 (2014).

\bibitem{dr29} H. Fujisaki:
EPL {\bf 28}, 623 (1994).

\bibitem{dr30} M. W. Kalinowski, M. Seweryński and L. Szymanowski:
JMP {\bf 24}, 375 (1983).

\bibitem{dr31} R. Contino and A. Gambassi:
JMP {\bf 44}, 570 (2003).

\bibitem{dr32} M. Dutsch, K. Fredenhagen, K. J. Keller and K. Rejzner3:
JMP {\bf 55}, 122303 (2014).

\bibitem{dr33} T. Nguyena:
JMP {\bf 57}, 092301 (2016).

\bibitem{dr34} J. Ben Geloun and R. Toriumi:
JMP {\bf 56}, 093503 (2015).

\bibitem{dr35} J. Ben Geloun and R. Toriumi:
J. Phys. A {\bf 45}, 374026 (2012).

\bibitem{dr36} B. Mutet, P. Grange and E. Werner:
J. Phys. A {\bf 45}, 315401 (2012).

\bibitem{dr37} M. C Abbott and P. Sundin:
J. Phys. A {\bf 45}, 025401 (2012).

\bibitem{dr38} T Fujihara et al.:
J. Phys. A {\bf 39}, 6371 (2008).

\bibitem{dr39} Silke Falk et al.:
J. Phys. A {\bf 43}, 035401 (2010).

\bibitem{dr40} Germán Rodrigo et al.:
J. Phys. G {\bf 25}, 1593 (1999).

\bibitem{dr41} B. M. Pimentel and J. L. Tomazelli:
J. Phys. G {\bf 20}, 845 (1994).

\bibitem{dr42}  A. Khare:
J. Phys. G {\bf 3}, 1019 (1977).

\bibitem{dr43}  J. C. D'Cruz:
J. Phys. G {\bf 1}, 151 (1975).

\bibitem{dr44}  R. Sepahv and S. Dadfar:
Nuc. Phys. A {\bf 960}, 36 (2017).

\bibitem{dr45}  J. V. Steele and R. J. Furnstahl:
Nuc. Phys. A {\bf 630}, 46 (1998).

\bibitem{dr46}  D. R. Phillips, S. R. Beane and T. D. Cohena:
Nuc. Phys. A {\bf 631}, 447 (1998).

\bibitem{dr47}  A. J. Stoddart and R. D. Viollier:
Nuc. Phys. A {\bf 532}, 657 (1991).

\bibitem{dr48}  E. Panzer:
Nuc. Phys. B {\bf 874}, 567 (2013).

\bibitem{dr49} R. N. Lee, A. V. Smirnov and and V. A. Smirnov:
Nuc. Phys. B {\bf 856}, 95 (2012).

\bibitem{dr50} A. P. Isaev:
Nuc. Phys. B {\bf 662}, 461 (2003).

\bibitem{dr51} J. M. Campbell, E. W. N. Glover and  D. J. Miller:
Nuc. Phys. B {\bf 498}, 397 (1997).

\bibitem{dr52} C. J. Yang, M. Grasso, X. Roca-Maza, G. Colo, and K. Moghrabi:
Phys. Rev. C {\bf 94}, 034311 (2016).

\bibitem{dr53} K. Moghrabi and M. Grasso:
Phys. Rev. C {\bf 86}, 044319 (2012).

\bibitem{dr54} D. R. Phillips, I. R. Afnan, and A. G. Henry-Edwards:
Phys. Rev. C {\bf 61}, 044002 (2000).

\bibitem{aa} A. Plastino, M. C. Rocca, Physica A 
{\bf 503}, 793 (2018), {\bf 497}, 310 (2018).

\bibitem{w1} C. G. Bollini, M. C. Rocca: arXiv:1012.4108, [hep-th].

\bibitem{w2} C. G. Bollini, M. C. Rocca: 
Int.J.Theor.Phys. {\bf 37} 2877 (1999). arXiv:hep-th/9807010.

\bibitem{w3} A. Wheeler and R. Feynman, Rev. Mod. Phys.
{\bf 17}, 157 (1945).

\bibitem{w4} A. Wheeler and R. Feynman, Rev. Mod. Phys.
{\bf 21}, 425 (1949).

\bibitem{tp7} I. M. Gel'fand and G. E. Shilov : ``Generalized
Functions'' {\bf Vol. 1}. Academic Press (1964).

\bibitem{tp17} S. Bochner:''Lectures on Fourier Integrals''.Princeton University
Press, Princeton, NJ, 1959.

\bibitem{tp13} I. S. Gradshteyn and I. M. Ryzhik : ``Table of Integrals,
Series and Products''. Academic Press, Inc (1980).

\end{thebibliography}
\end{document}